\documentclass[11pt,twoside]{article}
\usepackage{asp2010}
\usepackage{epsf}
\usepackage{psfig}

\resetcounters

\begin{document}

\title{Criss-cross mapping BD+30 3639: a new kinematic analysis technique}
\author{Wolfgang Steffen$^1$, Francisco Tamayo$^1$, and Nico Koning$^2$
\affil{$^1$Instituto de Astronom\'{\i}a, Universidad Nacional Aut\'onoma de M\'exico, C.P. 22860, Ensenada, M\'exico}
\affil{$^2$Department of Physics and Astronomy, University of Calgary, Calgary, Canada}}

\begin{abstract}

We present a new analysis of kinematic data of the young planetary nebula BD+30 3639. The data include spectroscopic long-slit and internal proper motion measurements. In this paper we also introduce a new type of mapping of kinematic proper motion data that we name ``criss-cross'' mapping. It basically consists of finding all points where extended proper motion vectors cross converge. From the crossing points a map is generated which helps to interpret the kinematic data. From the criss-cross mapping of BD+30 3639, we conclude that the kinematic center is approximately 0.5 arcsec off-set to the South-East from the central star. The mapping does also show evidence for a non-homologous expansion of the nebula that is consistent with a disturbance aligned with the bipolar molecular bullets.

\end{abstract}

\noindent {\bf Introduction} \\

\begin{figure}[!ht]
\plotone{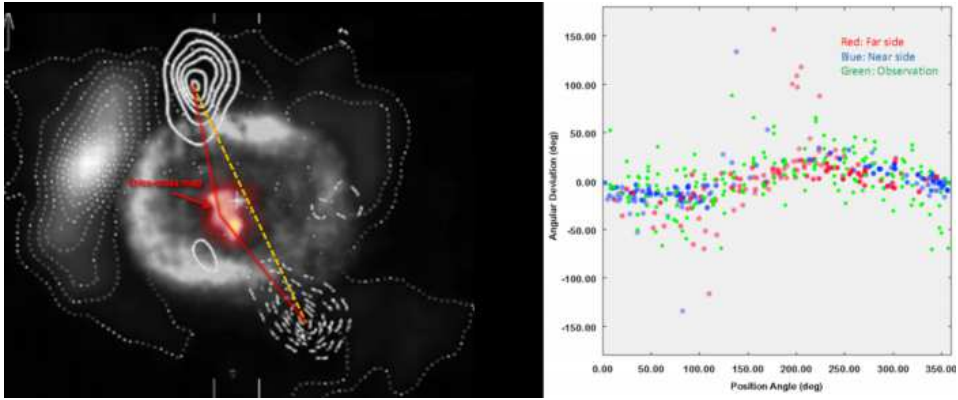}
\caption{On the left several types of maps have been combined: optical (grey, from LHB), H$_2$ (contoured grey, from Shupe et al., 1998), CO (contour, from Bachiller et al., 2000) and the criss-cross map (red, this work). On the right the angular deviation from the radial direction of the internal proper motion vectors is plotted against position angle. Data from LHB are in green and model values from this work are in red and blue.}
\label{steffen_fig1.fig}
\end{figure}

BD+30 3639 is a young planetary nebula that has been observed in detail with a variety of observational techniques. It shows extended emission in radio up to X-rays. The structure is, however, substantially different in the various wavelength regions. Some of the most striking features are as follows. The basic projected structure is a nearly rectangular ring, with some emission inside. The ring is not uniform along its perimeter, but shows reduced brightness especially in the south-western region. A fainter halo has been observed to go out to at least twice the distance of the ring (Harrington et al., 1997). The infrared continuum emission roughly traces the optical rectangle and its halo. Molecular hydrogen emission is distributed very unevenly in large clumps within the halo (Shupe et al., 1998). The molecular CO lines show a pair of high-speed bullets moving in opposite directions (Bachiller et al., 2000). Extended X-ray emission has been observed inside the optical rectangle with a brightness gradient going roughly from south-west to north-east (Kastner et al., 2000). Figure~\ref{steffen_fig1.fig}~(left) is a composite image that shows images taken in the different wavelength regions joint within a single frame.
Proper motion, combined with Doppler-velocity measurements and an accurate 3D model of the object may help improve the distance determination of the object. Li, Harrington \& Borkowski (2002, hereafter LHB) have obtained two HST narrow-band images that were observed with a separation of 5.663 years. They determined the expansion of the nebula along many angular sectors as well as local proper motion vectors of substructure at nearly 200 positions. With their measurements and an ellipsoidal model of BD+30 3639 they determine a distance of 1.2~kpc.

One of their results is that the expansion seems to be somewhat faster along position angles (PA) around 40$\deg$ and 220$\deg$. This coincides approximately with the position angles of the CO outflows. LHB concentrate on the variation of the magnitude of the proper motion vectors as a function of position angle and distance from the central star. In the present paper, in order to further improve the 3D model and the distance determination, we analyze their proper motion vectors with emphasis on the direction, i.e. their deviation from the radial direction, as a function of position angle and distance from the central star. We first describe a new method to visualize and analyze such data in the form of what we call ``criss-cross mapping''. A detailed 3D model and the distance determination will be presented in a separate paper. \\

  \noindent {\bf Criss-cross mapping} \\

In order to aid in the interpretation of current and future internal proper motion measurements in expanding nebulae we introduce ``criss-cross'' mapping. The basic idea is to emphasize regions where velocity vectors converge and intersect. Note that a radially expanding nebula will have all its velocity vectors intersect at the position of the central star. If there are systematic deviations from radial expansion, the intersection point might shift or be transformed into some extended pattern. Such a pattern might reveal helpful information. We therefore define the criss-cross mapping in its most basic form as follows: \\
\begin{figure}[!ht]
\plotone{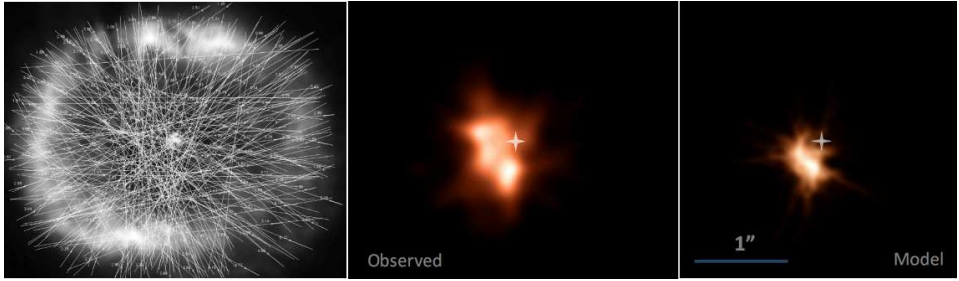}
\caption{Straight lines have been overlayed on all proper motion vectors of Fig.~3 from LHB (left). The resulting line image has then been convolved with a gaussian kernel to produce the criss-cross map in the middle, which is also shown in red in Figure~\ref{steffen_fig1.fig} (left) of this work. On the right is the corresponding criss-cross map from our {\em Shape} model.}
\label{steffen_fig2.fig}
\end{figure}
Replace every proper motion vector with a thin line that extends over the complete area covered by the nebula. Assign a finite constant brightness to every such line. Generate an image by adding together all lines. Finally, the result is convolved with a suitable kernel, like a gaussian with a width that is larger than the average separation between the vectors. In the regions where vectors converge the resulting image will increase in brightness and prominently reveal the regions where most velocity vectors meet. Details of the mapping theory and extensions to the scheme presented here will be published elsewhere (Steffen \& Koning, in prep.). We have implemented this simple procedure in the software {\em Shape}, which allows one to generate criss-cross maps from observations and models for immediate comparison.\\

  \noindent {\bf The direction of proper motion vectors in BD+30 3639} \\

From Figures 3 \& 4 of LHB we determined the PAs and deviation $\delta$ from the radial direction by direct measurement. We estimate the error in the measurement to be of the order of 2$\deg$ in PA and $\delta$. The distribution of $\delta$ as a function of PA is plotted in Fig.~\ref{steffen_fig1.fig}~(right). We find that the distribution is not random around zero, as would be expected for a radial expansion with some random measurement errors. Instead, the deviation from radial direction follows approximately a sinusoidal pattern.\\

  \noindent {\bf Modeling} \\

We have used the 3D morpho-kinematic modeling and reconstruction software {\it Shape} (Steffen et al., 2011) to reconstruct the 3D structure of BD+30 3639 based on the available imaging and internal proper motion from LHB and P-V diagrams from Bryce \& Mellema (1999).
For this initial reconstruction, we only used the [NII] data as a reference, since they are expected to be more thin\-shell like in structure, rather than span a significant range in distance for every given direction from the central star. Working with thin shells reduces ambiguities in the reconstructions.

In this paper, we concentrate on explaining the sinusoidal pattern in Fig.~\ref{steffen_fig1.fig}~(right). We found two quite different explanations. Steffen, Garc\'{\i}a-Segura \& Koning (2009, SGK) showed that the hydrodynamics of bipolar planetary nebulae is likely to have a non-homologous expansion with a poloidal velocity component. Since BD+30 3639 has a bipolar molecular ejection, we investigated whether the observations could be explained by a velocity field similar to those found by SGK. We find that the observed deviations from a radial velocity pattern can be explained if a cylindrical component is added to a homologous velocity field. Furthermore, the cylindrical component on the far side of the nebula is much smaller than on the near side. Alternatively, the observed velocity vectors correspond
mainly to structure on the near side. Although there is obscuration in the nebula, we find these solutions rather unlikely.

The second solution was found after introducing the new criss-cross mapping, which shows that the kinematic center is not located at the central star. Fig.~\ref{steffen_fig1.fig}~(left) shows the criss-cross map for BD+30 3639 superimposed on the observational composite image. There are well defined peaks at approx. 0.5~arcsec from the central star. This map shows that the kinematic center of expanding nebula is not located at the position of the central star. Figure \ref{steffen_fig2.fig}~(right) is a model map which includes a 0.5~arcsec shift of the velocity field in the direction as deduced from the observed criss-cross map. In addition to a homologous velocity component, there is a random noise in the velocity vector components of 4~km/s in each cartesian direction, as well as a cylindrical velocity component of 12~km/s along the direction of the molecular outflows. The cylindrical component is suppressed near the equatorial ring. The substructure of the observed and modeled criss-cross maps is similar. However, the model map lacks the northern peak that is present in the observed map. \\

\noindent {\bf Discussion and Conclusions} \\

The application to BD+30 3639 of our new criss-cross mapping technique leads us to conclude that the kinematic center is offset from the central star.  The lines connecting the molecular outflow with the central star and the peaks of the criss-cross map suggests the tails of the outflows might be directed towards the newly deduced kinematic center (Fig.~\ref{steffen_fig1.fig}). This conclusion does require confirmation, since the elliptical beam of the molecular map is approximately aligned with the direction between the southern outflow and the peak and structure of the criss-cross map. This problem does not, however, occur for the northern component, which shows a similar alignment with the criss-cross structure.

Reasons for the offset of the kinematic center could be motion of the central star within the nebula. There is, however, no evidence for that, since the star appears to be well centered on the optical image of the nebula. Another option is the presence of a secondary object that is responsible for the ejection of the bipolar molecular outflow. The distance of the object would be of 600 AU or more from the central star. The fact that such an object has not been detect so far sets strong constraints on its nature. The molecular outflow might have distorted the velocity field producing the observed offset and deviations from a homologous expansion.

\acknowledgements
This work has been supported by grants from CONACYT 49447 and UNAM PAPIIT IN100410. N.K. received additional support from the Natural Sciences, Alberta Ingenuity Fund and Engineering Research Council of Canada and from the Killam Trust.


\end{document}